\documentclass[aps,twocolumn,superscriptaddress,prd,10pt,nofootinbib]{revtex4-2}
\usepackage{graphicx}
\usepackage{amsmath}
\usepackage{amssymb}
\usepackage{amsfonts}
\usepackage{color}
\usepackage{dsfont}
\usepackage{soul}
\usepackage{hyperref}

\usepackage{tikz}
\usetikzlibrary{positioning,shapes}

\usepackage{tikz,xcolor}
\definecolor{lime}{HTML}{A6CE39}
\DeclareRobustCommand{\orcidicon}{%
	\begin{tikzpicture}
	\draw[lime, fill=lime] (0,0) 
	circle [radius=0.16] 
	node[white] {{\fontfamily{qag}\selectfont \tiny ID}};
	\draw[white, fill=white] (-0.0625,0.095) 
	circle [radius=0.007];
	\end{tikzpicture}
	\hspace{-2mm}
}
\foreach \x in {A, ..., Z}{%
	\expandafter\xdef\csname orcid\x\endcsname{\noexpand\href{https://orcid.org/\csname orcidauthor\x\endcsname}{\noexpand\orcidicon}}
}

\begin{document}

\title{Gravitational redshift induces quantum interference}
\date{\today}

\author{David Edward Bruschi\orcidA{}}
\address{Institute for Quantum Computing Analytics (PGI-12), Forschungszentrum J\"ulich, 52425 J\"ulich, Germany}
\email{david.edward.bruschi@posteo.net}
\author{Andreas Wolfgang Schell\orcidB{}}
\address{Institut f\"ur Festk\"orperphysik, Leibniz Universit\"at Hannover, 30167 Hannover, Germany}
\address{Physikalisch-Technische Bundesanstalt, 38116 Braunschweig, Germany}

\begin{abstract}
We use quantum field theory in curved spacetime to show that gravitational redshift induces a unitary transformation on the quantum state of propagating photons. We find that the transformation is a mode-mixing operation, and we devise a protocol that exploits gravity to induce a Hong-Ou-Mandel-like interference effect on the state of two photons. We discuss how the results of this work can provide a demonstration of quantum field theory in curved spacetime.
\end{abstract}

\maketitle

\section{Introduction}
Gravitational redshift is a trademark prediction of general relativity \cite{Misner:Thorne:1973,Wilhelm:Bhola:2014}. Photons initially prepared with a given frequency by the sender travel through curved spacetime and are detected with a different frequency by the receiver. This effect, which can be successfully explained by general relativity alone, has been tested and measured using a plethora of different setups \cite{Pound:Rebka:1959,Mueller:Wold:2006,Chou:Hume:2010,Mueller:Peters:2010,Clifford:2014,Litvinov:Rudenko:2018,DiPumpo:Ufrecht:2021}, and can even be exploited for different tasks \cite{Bruschi:Ralph:2014,Bruschi:Datta:2014,Kohlrus:Bruschi:2017}. 

In recent years, renewed attention to the overlap of quantum mechanics and relativity has been fuelled by developments in quantum information theory \cite{Nielsen:Chuang:2010}. Many experimental and theoretical proposals have been put forward to exploit inherent features of quantum systems, such as entanglement, to measure gravitationally induced decoherence of a quantum state \cite{Bassi:Grossart:2017,Howl:Penrose:2019}, test the quantum nature of gravity with tabletop experiments \cite{Carney:Stamp:2019}, exploit interferometric setups to test gravitationally-induced effects on the interferometric visibility \cite{Williams:Chiow:2016,Tino:2021}, understand quantum clocks within relativistic settings \cite{Smith:Ahmadi:2019,Smith:Ahmadi:2020}, and investigate the interplay between gravity and quantum correlations present in the state of a quantum system \cite{Bruschi:Wilhelm:2020}. 
Gravitational redshift often plays a key role in this field of research. Therefore, it is important to understand if it can be implemented as a quantum operation. 

In this work we ask the question: \textit{is gravitational redshift implemented as a unitary transformation in a physical process? If it is, can we characterize such transformation?} To answer this question we use quantum field theory in curved spacetime to model a pulse of light that propagates on a classical curved background. We focus in particular on static spacetimes, where a gravitational redshift can be defined meaningfully between a sender and a receiver who are at rest \cite{Wilhelm:Bhola:2014}. 

\begin{figure}[ht!]
    \centering
    \includegraphics[width=0.9\linewidth]{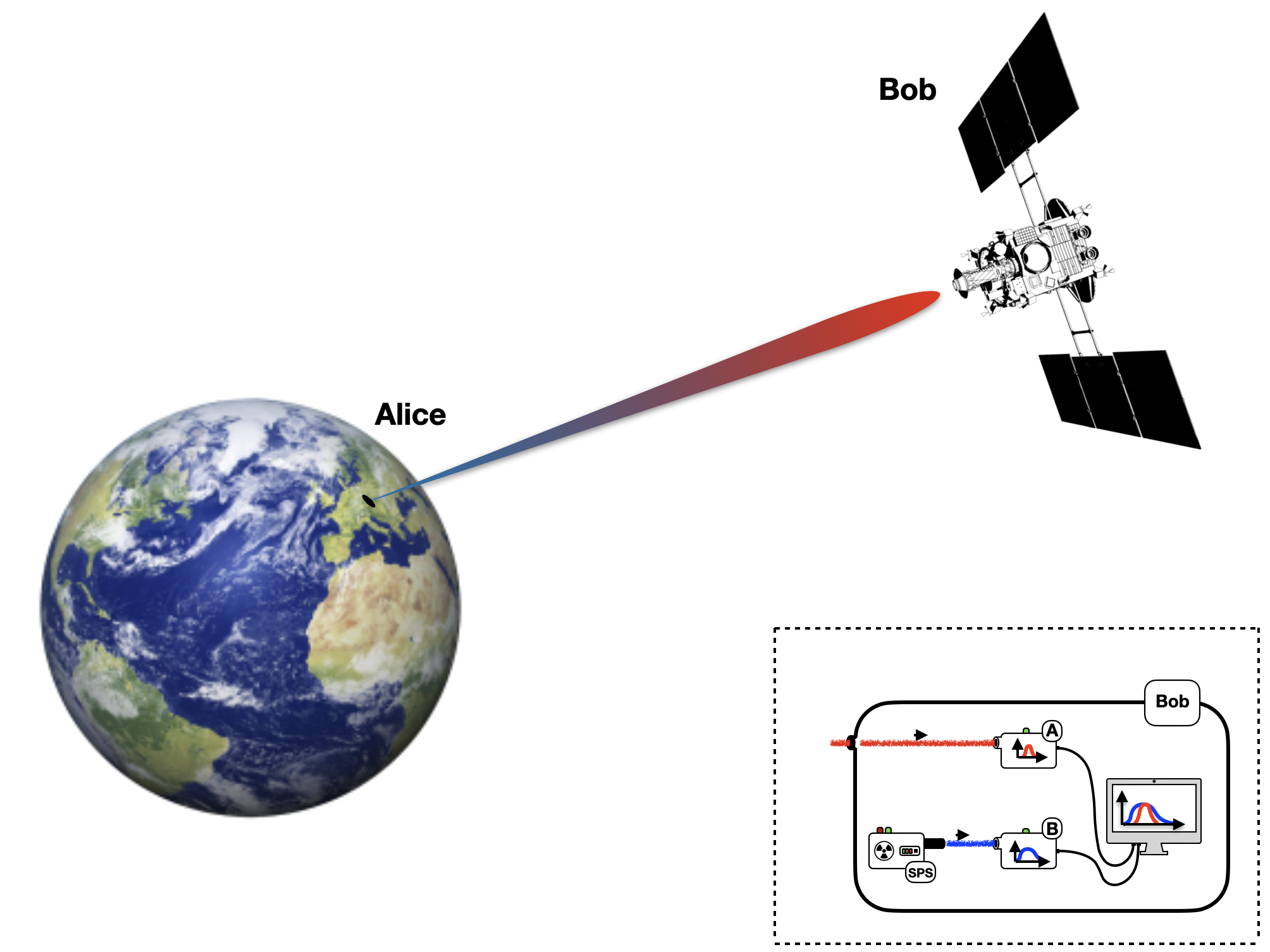}
    \caption{Alice and Bob agree on a frequency profile of photons that they will exchange. Alice sends a photon, or pulses of light, to Bob who will, in general, receive a different frequency profile due to gravitational redshift. Alice's photon is detected ({\color{red} red}) by Bob at photodetector A, which he can compare \textit{locally} with his photons ({\color{blue} blue}) \textit{identical} to the expected one. Discrepancies indicate that the input photon has undergone a transformation that Bob wishes to characterize.}
    \label{fig1}
\end{figure}

\section{Background tools}

\subsection{Quantum fields in curved spacetime}
Let us consider, without loss of generality, a massless scalar quantum field $\hat{\phi}(x^\mu)$ propagating on classical (curved) $3+1$ background with coordinates $x^\mu$ and metric $g_{\mu\nu}$.\footnote{An introduction to quantum field theory in curved spacetime is left to standard references \cite{Birrell:Davies:1982}. The metric has signature $(-,+,+,+)$. We use Einstein's summation convention and natural units $c=\hbar=G_\text{N}=1$ unless explicitly stated. We work in the Heisenberg picture.}  The classical field $\phi(x^\mu)$ will satisfy the Klein-Gordon equation 
\begin{equation}\label{equation:one}
\left((\sqrt{-g})^{-1}\partial_\mu g^{\mu\nu} \sqrt{-g} \partial_\nu\right)\phi(x^\mu)=0.
\end{equation}
Finding solutions to Equation~\eqref{equation:one} is very difficult since, in a general spacetime, there is no preferred notion of time \cite{Misner:Thorne:1973,Birrell:Davies:1982}. When a notion of time exists, for example the spacetime has a global timelike Killing vector field $\partial_t$, it is possible to meaningfully foliate the spacetime in spacelike hypersurfaces orthogonal to $\partial_t$ and solve the Klein-Gordon equation, to finally obtain upon quantization $\hat{\phi}(x^\mu)=\int d^3k\,[\phi_{\boldsymbol{k}}(x^\mu)\,\hat{a}_{\boldsymbol{k}}+\phi_{\boldsymbol{k}}^*(x^\mu)\,\hat{a}_{\boldsymbol{k}}^\dag]$. The mode solutions $\phi_{\boldsymbol{k}}(x^\mu)$ are labelled by $\boldsymbol{k}\equiv(k_x,k_y,k_z)$, satisfy $\square\phi_{\boldsymbol{k}}(x^\mu)=0$ (which is a short-hand notation for Equation~\eqref{equation:one}), are normalized by $(\phi_{\boldsymbol{k}},\phi_{\boldsymbol{k}'})=\delta^{3}(\boldsymbol{k}-\boldsymbol{k}')$ given the appropriate inner product $(\cdot,\cdot)$, and the annihilation and creation operators satisfy $[\hat{a}_{\boldsymbol{k}},\hat{a}_{\boldsymbol{k}'}^\dag]=\delta^{3}(\boldsymbol{k}-\boldsymbol{k}')$, while all others vanish. The mode solutions also satisfy $i\,\partial_t\phi_{\boldsymbol{k}}(x^\mu)=\omega_{\boldsymbol{k}}\,\phi_{\boldsymbol{k}}(x^\mu)$, which guarantees a consistent notion of particle in time. In general, $\omega_{\boldsymbol{k}}$ is a function of $\boldsymbol{k}$ and, for example, in flat spacetime one has $\omega_{\boldsymbol{k}}=|\boldsymbol{k}|$.

\subsection{Modelling a realistic photon}
We have chosen to use a massless scalar field, which can be employed to model one polarization of the electromagnetic field in the regimes considered here \cite{Srednicki:2007}. Photons defined by the operators $\hat{a}_{\boldsymbol{k}}$ are ideal, in the sense that their spatial support occupies the whole spacetime. Consequently, they cannot be employed to discuss concrete physical effects since the modes $\phi_{\boldsymbol{k}}(x^\mu)$ are normalized through a Dirac-deltas. 

A realistic photon, on the other hand, is characterized by a finite spatial extension and frequency bandwidth instead of an (infinitely) sharp frequency. We assume that we can discard all effects due to the extension of the photon along directions that are orthogonal to that of propagation, and that these can be taken into account separately \cite{Exirifard:Culf:2021,Exirifard:Karimi:2022}. A photon operator is therefore constructed as  
\begin{equation}
\hat{A}_{\omega_0}:=\int_0^\infty d\omega F_{\omega_0}(\omega/\sigma)\,\hat{a}_\omega,
\end{equation}
where the (complex) function $ F_{\omega_0}(\omega/\sigma)$ determines the frequency profile. This function is labelled by the peak frequency $\omega_0$, it has an overall characteristic size $\sigma$, and it is normalized by $\langle F_{\omega_0},F_{\omega_0}\rangle=1$, where we define $\langle F,G\rangle:=\int_0^\infty d\omega F^*(\omega)G(\omega)$ for later convenience. It is immediate to check that $[\hat{A}_{\omega_0},\hat{A}_{\omega_0}^\dag]=1$, which therefore guarantees that $\hat{A}_{\omega_0}^\dag$ generates properly normalized photonic states. We note that the Hilbert space $\mathcal{H}$ of the (scalar) photon is infinite dimensional and therefore we need to introduce the set of functions $F_{\underline{\lambda}}$ determined by a set of parameters $\underline{\lambda}$ such that, together with $F_{\omega_0}$, they form an orthonormal basis. In practice this means that $\langle F_{\omega_0},F_{\underline{\lambda}}\rangle=0$ for all $\underline{\lambda}$, while $\langle F_{\underline{\lambda}},F_{\underline{\lambda}'}\rangle=\delta(\underline{\lambda}-\underline{\lambda}')$. Here $\underline{\lambda}$ is, in principle, a collection of discrete and continuous indices. Operators can then be defined as $\hat{A}_{\underline{\lambda}}:=\int_0^\infty d\omega\,F_{\underline{\lambda}}(\omega)\,\hat{a}_\omega$ and therefore $[\hat{A}_{\omega_0},\hat{A}^\dag_{\underline{\lambda}}]=0$. In this work we do not necessitate an explicit construction of the set $\{F_{\underline{\lambda}}\}$.

\subsection{Gravitational redshift}
Gravitational redshift is a key prediction of general relativity, which lacks a conclusive explanation \cite{Okun:2000,Wilhelm:Bhola:2014}. It remains unclear if it is a fundamental effect witnessed by the photons, or a consequence of the effects of gravity on local measuring devices. In the second case, gravitational redshift is not a ``change in frequency of the photon'', but a mismatch in the frequencies of the constituents forming, for example, the detecting devices of the sender and receiver respectively. Here we take the approach that a frequency is what a (localized) observer measures with his (local) clock. With this in mind, we assume that Alice and Bob are stationary and therefore don't have to correct for additional effects due to relative motion, i.e., for Doppler-like effects. Alice measures proper time $\tau_{\textrm{A}}$ locally at A using her clock, while Bob measures proper time $\tau_{\textrm{B}}$ locally at B using his. We then recall that the relation between the frequency $\omega_{\textrm{A}}$ prepared by Alice at position A, and the frequency $\omega_{\textrm{B}}$ received by Bob at location $B$, is
\begin{align}\label{general:redshift}
\chi^2:=\frac{\omega_{\textrm{B}}}{\omega_{\textrm{A}}}=\frac{k_\mu\,u^\mu_{\textrm{B}}}{k_\mu\,u^\mu_{\textrm{A}}},
\end{align}
where $k_\mu$ is the tangent vector to the (affinely parametrized) null geodesic followed by the photon, $u^\mu_{\textrm{B}}$ is the Alice's four velocity and $u^\mu_{\textrm{A}}$ is Bob's four velocity \cite{Wald:1984}. It is understood that $k_\mu\,u^\mu_{\textrm{A}}$ and $k_\mu\,u^\mu_{\textrm{B}}$ are calculated at Alice's and Bob's positions respectively. The nonnegative parameter $\chi$ has been introduced for notational convenience and is central to this work.
While this relation is central to our work, we do not join the debate on the interpretation of the redshift presented above. We note, however, that the effects found here are witnessed locally by the observers when they measure the quantum states of light.

\section{Gravitational redshift of realistic photons}

\subsection{Gravitational redshift of photon operators}
Alice and Bob wish to determine how gravitational redshift affects photons. Alice sends a photon to Bob, who will detect a gravitational redshift within the incoming photon, i.e., each sharp frequency $\omega'$ as measured \textit{locally} by his clock will not coincide with the sharp frequency $\omega$ of the sent photon. The scheme is depicted in Figure~\ref{fig1}. 
As far as Bob is concerned, i.e., from the perspective of his laboratory, the expected photon has changed and he can study the properties of the transformation involved, \textit{irrespective} of where the incoming photon has originated or which specific physical process it has undergone. Therefore, Bob can assign a channel to the process that affected the incoming photon, and seek for its properties. 

Bob starts by assuming that there is a transformation $T(\chi):\omega\rightarrow\chi^2\omega$ on \textit{each} sharp frequency $\omega$. He then \textit{looks for a unitary transformation $\hat{U}(\chi)$ that implements $T(\chi)$ through}
\begin{align}\label{unitary:transformation}
\hat{a}_{\omega'}=\hat{U}^\dag(\chi)\,\hat{a}_{\omega}\,\hat{U}(\chi)=\hat{a}_{\chi^2\omega}
\end{align}
for all $\chi$, where $\hat{U}^\dag(\chi)\hat{U}(\chi)=\mathds{1}$. 

Assuming that the transformation \eqref{unitary:transformation} holds, it is easy to use the explicit expression for $\hat{A}_{\omega_0}$, the fact that $[\hat{a}_{\chi^2\omega},\hat{a}_{\chi^2\omega'}^\dag]=\delta(\chi^2\omega-\chi^2\omega')$, and $\delta(f(x))=\sum_n\delta(x-x_{0,n})/|f'(x_{0,n})|$, where $x_{0,n}$ are the zeros of the function $f(x)$, to show that 
\begin{equation}
1=\hat{U}^\dag(\chi)\hat{U}(\chi)=\hat{U}^\dag(\chi)[\hat{A}_{\omega_0},\hat{A}_{\omega_0}^\dag]\hat{U}(\chi)=\frac{1}{\chi^2}.
\end{equation}
This equation can be satisfied only when $\chi=1$, that is, for the trivial case of no redshift. Clearly, this cannot happen in general as can be seen from \eqref{general:redshift}. Therefore, we conclude that gravitational redshift in the form of a linear shift of the spectrum of sharp frequencies \textit{cannot} be obtained as the result of a unitary operation on the field modes $\{\hat{a}_\omega\}$ alone.\footnote{Note that if \eqref{unitary:transformation} were replaced by $\hat{a}_{\omega'}=\hat{U}^\dag(\chi)\,\hat{a}_{\omega}\,\hat{U}(\chi)=\chi\hat{a}_{\chi^2\omega}$, the commutation relations would be persevered.} This result corroborates the claim that the gravitational redshift is \textit{not} simply a shift in the sharp frequencies of the photons for all frequencies of the spectrum. 

\subsection{Quantum modelling of gravitational redshift}
We now ask a more refined version of the question posed above: \textit{how is the transformation $T(\chi)$ implemented by a unitary operator when acting on realistic photons?} To answer this question, we start by noting that Bob will describe the received photon as $\hat{A}_{\omega_0'}=\int_0^\infty d\omega\,F'_{\omega_0'}(\omega/\sigma')\,\hat{a}_\omega$ as a function of the \textit{local} frequency $\omega$ as measured in his laboratory, while the expected photon has the expression $\hat{A}_{\omega_0}=\int_0^\infty d\omega\,F_{\omega_0}(\omega/\sigma)\,\hat{a}_\omega$. Bob then notices that each sharp frequency $\omega$ that appears in the definition of $\hat{A}_{\omega_0}$ transforms by $T(\chi):\omega\rightarrow\chi^2\omega$, see \cite{Bruschi:Ralph:2014}. This means that 
\begin{equation}
\int_0^\infty d\omega\,F_{\omega_0}(\omega/\sigma)\,\hat{a}_\omega\overset{T(\chi)}{\rightarrow}\chi^2\int_0^\infty d\omega\,F_{\omega_0}(\chi^2\omega/\sigma)\,\hat{a}_{\chi^2\omega}. 
\end{equation}
He can then identify the function $F'_{\omega_0'}(\omega/\sigma')\equiv\chi\,F_{\omega_0}(\chi^2\omega/\sigma)=\chi\,F_{\omega_0/\chi^2}(\omega/(\sigma/\chi^2))$, where $\omega_0'=\omega_0/\chi^2$ and $\sigma'=\sigma/\chi^2$, which represents a well defined physical photon in the sense that $\int_0^\infty d\omega|F'_{\omega_0'}(\omega/\sigma')|^2=1$. He is left with introducing the operator $\hat{a}_{\omega}':=\chi\hat{a}_{\chi^2\omega}$, which has well defined canonical commutation relations $[\hat{a}_{\omega}',\hat{a}_{\omega}^{\dag\prime}]=\delta(\omega-\omega')$. We now note that the fact that $\hat{a}_{\omega}',\hat{a}_{\omega}^{\dag\prime}$ and $\hat{a}_{\omega},\hat{a}_{\omega}^\dag$ have identical commutation relations \textit{for the same  frequencies}, and the fact that $\int d\omega\,\hbar\omega\,\hat{a}_{\omega}^{\dag\prime} \hat{a}_{\omega}^{\prime}| 1_{\omega}'\rangle=\hbar\omega|1_\omega'\rangle$, where $|1_\omega'\rangle:=\hat{a}_{\omega}^{\dag\prime}|0\rangle$, imply that Bob cannot distinguish locally between $\hat{a}_{\omega}$ and $\hat{a}_{\omega}'$, and he therefore identifies $\hat{a}_{\omega}'\equiv\hat{a}_{\omega}$. 
Bob therefore can assume that the following unitary transformation has occurred
 \begin{align}\label{arbitrary:transformation:extended}
\hat{A}_{\omega_0'}=&\hat{U}^\dag(\chi)\,\hat{A}_{\omega_0}\,\hat{U}(\chi)=\int_0^\infty d\omega\,F'_{\omega_0'}(\omega/\sigma')\hat{a}_{\omega},
\end{align}
with the relation $F'_{\omega_0'}(\omega/\sigma')\equiv \chi F_{\omega_0}(\chi^2\omega/\sigma)$.
Notice that the transformation \eqref{arbitrary:transformation:extended} applies appropriately to \textit{all} of the photon operators $\{\hat{A}_{\omega_0},\hat{A}_{\underline{\lambda}}\}$ and implies, equivalently, that there is a canonical transformation of the bases $\{F_{\omega_0},F_{\underline{\lambda}}\}$ and $\{F'_{\omega_0'},F'_{\underline{\lambda}'}\}$ of mode functions. 

We can collect all field operators $\{\hat{A}_{\omega_0},\hat{A}_{\underline{\lambda}}\}$ in the vector $\hat{\mathbb{X}}:=(\hat{A}_{\omega_0},\hat{A}_{\underline{\lambda}_1},\ldots)^{\textrm{Tp}}$. Then, the transformation \eqref{arbitrary:transformation:extended} implies that there exists a unitary matrix $\boldsymbol{U}(\chi)$ such that
 \begin{align}\label{arbitrary:transformation:representation}
\hat{\mathbb{X}}':=\hat{U}^\dag(\chi)\,\hat{\mathbb{X}}\,\hat{U}(\chi)\equiv\boldsymbol{U}(\chi)\,\hat{\mathbb{X}}.
\end{align}
This transformation is known in quantum optics as a \textit{mode-mixer} \cite{Scully:Zubairy:1997}, and it is a particular case of a symplectic transformation \cite{Adesso:Ragy:2014}. Note that, if we choose $N$ modes to study, there are $N(N+1)$ independent overlaps of the form $\langle F_n,F_m\rangle$ (including the modulus and the phase). The overlaps with $F_{\perp}$ are uniquely fixed this way as well. A transformation of the form \eqref{arbitrary:transformation:representation}, on the other hand, is determined by the $(N+1)\times(N+1)$ unitary matrix $\boldsymbol{U}(\chi)$ that mixes the $N$ chosen modes with the orthogonal complement $F_{\perp}$. Therefore, the independent angles that define $\boldsymbol{U}(\chi)$ are $(N+1)N/2$, and there are also $(N+1)N/2$ independent phases. Since the degrees of freedom match in number, it is a well posed operation to identify the angles of $\boldsymbol{U}(\chi)$ through the \textit{independent} overlaps $|\langle F_n,F_m\rangle|$, and the phases of $\boldsymbol{U}(\chi)$ with $\textrm{arg}(\langle F_n,F_m\rangle)$, as also noted in the literature \cite{Menessen:Jones:2017}.

\section{Gravitational-redshift-induced interference}

\subsection{Gravitationally-induced tritter}
Let us focus on the case where we select two \textit{different} commuting photon operators $\hat{A}_{\omega_0}$ and $\hat{A}_{\tilde{\omega}_0}$, and let us consider the transformed modes $\hat{A}_{\omega_0'}$ and $\hat{A}_{\tilde{\omega}_0'}$. We then define the vector $\hat{\mathbb{X}}:=(\hat{A}_{\omega_0},\hat{A}_{\tilde{\omega}_0},\hat{A}_{\perp})^{\textrm{Tp}}$, where the operator $\hat{A}_{\perp}:=\sum_{\underline{\lambda}} \alpha_{\underline{\lambda}}\,\hat{A}_{\underline{\lambda}}$ collects all of the operators orthogonal to the two chosen ones, we have  $\sum_{\underline{\lambda}} |\alpha_{\underline{\lambda}}|^2=1$, and $\hat{A}_{\perp'}:=\hat{U}^\dag(\chi)\,\hat{A}_{\perp}\,\hat{U}(\chi)$. The general transformation \eqref{arbitrary:transformation:representation} is therefore defined by the symplectic representation of the product of three beam-splitting operations of the form $\exp[\theta(e^{i\varphi_\theta}\hat{A}_{\omega_0}\hat{A}_\perp^\dag-e^{-i\varphi_\theta}\hat{A}_{\omega_0}^\dag\hat{A}_\perp)]$, $\exp[\psi(e^{i\varphi_\psi}\hat{A}_{\tilde{\omega}_0}\hat{A}_\perp^\dag-e^{-i\varphi_\psi}\hat{A}_{\tilde{\omega}_0}^\dag\hat{A}_\perp)]$ and $\exp[\phi(e^{i\varphi_\phi}\hat{A}_{\omega_0}\hat{A}_{\tilde{\omega}_0}^\dag-e^{-i\varphi_\phi}\hat{A}_{\omega_0}^\dag\hat{A}_{\tilde{\omega}_0})]$. We report the explicit result for $\varphi_\theta=\varphi_\phi=\varphi_\psi=0$, and note that the phases can be restored when necessary. We have
\begin{align}\label{beam:tritter}
\boldsymbol{U}
\equiv
\begin{pmatrix}
\textrm{c}_\theta\,\textrm{c}_\phi\ & -\textrm{c}_\theta\,\textrm{s}_\phi\,\textrm{c}_\psi -\textrm{s}_\theta\textrm{s}_\psi & -\textrm{c}_\theta\,\textrm{s}_\phi\,\textrm{s}_\psi +\textrm{s}_\theta\textrm{c}_\psi\\
\textrm{s}_\phi & \textrm{c}_\phi\,\textrm{c}_\psi & \textrm{c}_\phi\,\textrm{s}_\psi\\
-\textrm{s}_\theta\,\textrm{c}_\phi & \textrm{s}_\theta\,\textrm{s}_\phi\,\textrm{c}_\psi -\textrm{c}_\theta\textrm{s}_\psi & \textrm{s}_\theta\,\textrm{s}_\phi\,\textrm{s}_\psi +\textrm{c}_\theta\textrm{c}_\psi
\end{pmatrix},
\end{align}
where we introduce $\textrm{s}_{\vartheta}:=\sin\vartheta$ and $\textrm{c}_{\vartheta}:=\cos\vartheta$ for ease of presentation. The transformation \eqref{beam:tritter} is known in quantum optics as a \textit{three-wave mode-mixer}, or \textit{tritter} \cite{Scully:Zubairy:1997,Menessen:Jones:2017}.

Most importantly, the angles $\theta$, $\phi$ and $\psi$ are functions of the redshift $\chi$ and are defined through 
\begin{align}\label{tritter:angles}
\cos\theta\,\cos\phi\equiv&|\langle F'_{\omega_0'},F_{\omega_0}\rangle|,\nonumber\\
\cos\phi\,\cos\psi\equiv&|\langle F'_{\tilde{\omega}_0'},F_{\tilde{\omega}_0}\rangle|,\nonumber\\
\sin\phi\equiv&|\langle F'_{\tilde{\omega}_0'},F_{\omega_0}\rangle|.
\end{align}
We expect that $\theta$, $\phi$ and $\psi$, in the redshift regime $\chi\geq1$, take values between $\theta=\phi=\psi=0$ (i.e., perfect overlap), and $\theta=\psi=\pi/2$, $\phi=0$ (complete mismatch). An analogous analysis can be done for the blueshift regime $0\leq\chi<1$.

\subsection{Gravity-induced quantum interference}
\begin{figure}[ht!]
    \centering
    \includegraphics[width=0.9\linewidth]{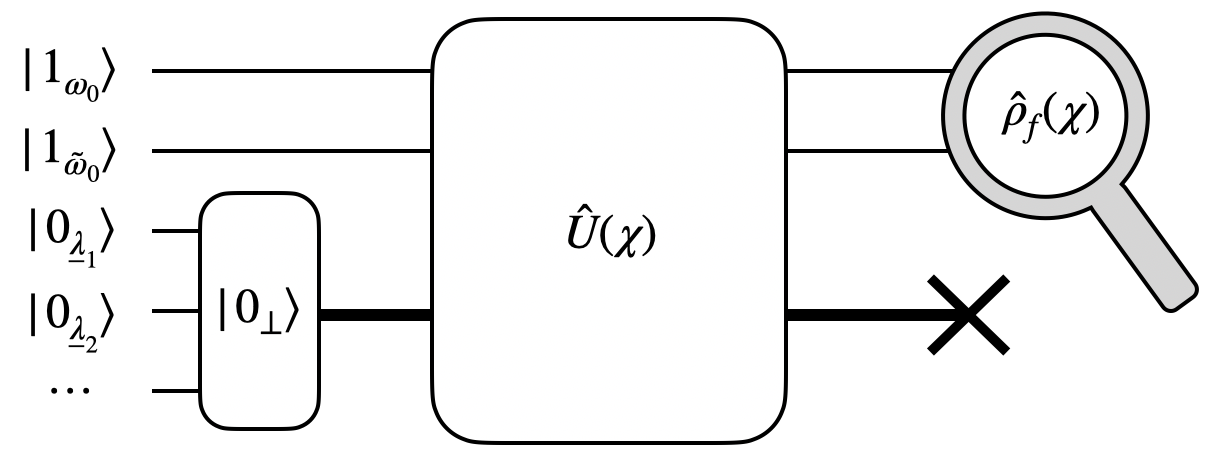}
    \caption{Alice sends a two-photon $|1_{\omega_0}1_{\tilde{\omega}_0}\rangle$ of modes $\hat{A}_{\omega_0}$ and $\hat{A}_{\tilde{\omega}_0}$ to Bob. The gravitational redshift effectively mode-mixes the state through the unitary operation $\hat{U}(\chi)$ defined in \eqref{beam:tritter} into components $\hat{A}_{\omega_0}$, $\hat{A}_{\tilde{\omega}_0}$ and $\hat{A}_\perp$. Bob then measures the reduced state of modes $\hat{A}_{\omega_0}$ and $\hat{A}_{\tilde{\omega}_0}$, which is now entangled.}
    \label{fig2}
\end{figure}

Here we describe a photon-exchange task between Alice and Bob that exploits the transformation \eqref{beam:tritter} to show that gravity induces quantum interference of photonic states. It is depicted using a circuit implementation language in Figure~\ref{fig2}.

The steps required to perform the task read as follows:
\begin{itemize}
\setlength\itemsep{-0.35em}
\item[i.] Alice prepares a two-photon \textit{separable} state $|\Psi_0\rangle:=|1_{\omega_0}1_{\tilde{\omega}_0}0\rangle$ and sends it to Bob, who receives it as $|\Psi\rangle:=|1_{\omega_0'}1_{\tilde{\omega}_0'}0\rangle$. Introducing the notation $|nmp\rangle:=\frac{(\hat{A}^{\dag}_{\omega_0})^n}{\sqrt{n!}}\frac{(\hat{A}^{\dag}_{\tilde{\omega}_0})^m}{\sqrt{m!}}\frac{(\hat{A}^{\dag}_\perp)^p}{\sqrt{p!}}|0\rangle$, it is immediate to verify that Bob's state reads \textit{locally} as
\begin{align}\label{general:tritter:state:Bob}
|\Psi\rangle=&\sqrt{2}\left[U_{13}U_{23}|002\rangle+U_{12}U_{22}|020\rangle+U_{11}U_{21}|200\rangle\right]\nonumber\\
+&(U_{13}U_{22}+U_{12}U_{23})|011\rangle+(U_{11}U_{22}+U_{12}U_{21})|110\rangle\nonumber\\
+&(U_{11}U_{23}+U_{13}U_{21})|101\rangle.
\end{align}
Here $U_{ab}$ are the coefficients of the matrix \eqref{beam:tritter}.
\item[ii.] The final state $\hat{\rho}_{\textrm{f}}(\chi)$ of the modes $\hat{A}_{\omega_0}$ and $\hat{A}_{\tilde{\omega}_0}$ in Bob's laboratory is obtained by tracing \eqref{general:tritter:state:Bob} over the unovserbed subsystem degrees of freedom $\hat{A}_\perp$, and it is easy to compute but gives a cumbersome expression. We give its generic form here
\begin{align}\label{general:reduced:tritter:state:Bob}
\hat{\rho}_{\textrm{f}}(\chi)=&\rho_{0000}|00\rangle\langle00|+\rho_{0202}|02\rangle\langle02|+\rho_{2020}|20\rangle\langle20|\nonumber\\
+&\rho_{1010}|10\rangle\langle10|+\rho_{0101}|01\rangle\langle01|+\rho_{1111}|11\rangle\langle11|\nonumber\\
+&\rho_{2011}|20\rangle\langle11|+\rho_{0211}|02\rangle\langle11|\nonumber\\
+&\rho_{2001}|20\rangle\langle02|+\rho_{1001}|10\rangle\langle01|+\textrm{h.c.}
\end{align}
The coefficients $\rho_{nmpq}$ can be obtained in terms of the matrix elements $U_{ab}$ with simple algebra. We avoid printing them here to improve clarity of presentation.
\end{itemize}
We note here that it is possible to have all terms in \eqref{general:reduced:tritter:state:Bob} that include a $|11\rangle$ contribution to vanish with either $\rho_{0202}$ or $\rho_{2020}$ remaining nonzero. It is sufficient that either $U_{11}=U_{21}=0$ while $U_{12},U_{22}$ are both nonzero, or viceversa. In the first case we obtain the fully mixed state $\hat{\rho}_{\textrm{f}}(\chi)=2|U_{13}U_{23}|^2|00\rangle\langle00|+2|U_{12}U_{22}|^2|02\rangle\langle02|+|U_{13}U_{22}+U_{12}U_{23}|^2|01\rangle\langle01|$. The other case can be obtained in a similar fashion.

More importantly, however, is the case when $\rho_{1111}=0$, but $\rho_{0202}\neq0$ and $\rho_{2020}\neq0$ \textit{at the same time}. This requires us to assume that $|U_{11}U_{22}+U_{12}U_{21}|=|c_\theta (c_\phi^2-s_\phi^2)c_\psi-s_\theta s_\phi s_\psi|=0$, which can occur given the freedom in choice of the initial modes. In this case, all terms in \eqref{general:reduced:tritter:state:Bob} with $|11\rangle$ vanish, and we are left with a state that exhibits Hong-Ou-Mandel-like interference \cite{Hong:Ou:1987,Bouchard:Sit:2021}. This is a genuine quantum effect due to gravity.

We can finally verify if the state \eqref{general:reduced:tritter:state:Bob} is entangled. This requires the partial transpose $\hat{\rho}^{\textrm{pt}}_{\textrm{f}}(\chi)$ (with respect, say, of the second mode) of the state, and the use of the \textit{negativity} $\mathcal{N}(\hat{\rho}_{\textrm{f}}(\chi)):=\max\{0,1/2\sum_{\lambda<0}(|\lambda|-\lambda)\}$, where $\lambda$ are the eigenvalues of $\hat{\rho}^{\textrm{pt}}_{\textrm{f}}(\chi)$. If the negativity is nonzero, the PPT criterion guarantees that the state is entangled \cite{Peres:1996}. We can only find explicitly two negative eigenvalues of the partial transpose, which are sufficient for the detection. In fact, some algebra gives us 
 \begin{align}\label{Gaussian:profile:example}
\mathcal{N}(\hat{\rho}_{\textrm{f}}(\chi))\geq&\frac{1}{2}\left[\sqrt{\rho_{0101}^2+4|\rho_{0211}|^2}-\rho_{0101}\right]\nonumber\\
&+\frac{1}{2}\left[\sqrt{\rho_{1010}^2+4|\rho_{2011}|^2}-\rho_{1010}\right],
\end{align}
which is greater than $0$ for values at least one of $\rho_{0211}$ or $\rho_{2011}$ greater than zero. When this occurs, we conclude that gravitational redshift has entangled the state. Note that in the case where $\rho_{1111}=0$ we also have $\rho_{0211}=\rho_{2011}=0$, which implies that the right-hand side of \eqref{Gaussian:profile:example} also vanishes. In this case, in order to detect entanglement we need to compute the other negative eigenvalues either analytically of numerically. This poses no conceptual difficulty, and can be therefore done when required.

\section{Considerations and Applications}
We now proceed and offer a few considerations regarding the formalism presented and used in the present work, as well as the predictions that we have put forward. We also comment on potential applications.
 
Our results depend on the validity of quantum field theory in curved spacetime. Therefore, testing the predictions of this work, such as the validity of the transformation \eqref{beam:tritter} for different redshifts $\chi$, i.e., different configurations of the Alice-Bob positioning, can be used to test the theory. In particular, it is possible to employ the protocol described above to verify when the state \eqref{general:reduced:tritter:state:Bob} can be obtained in the first place, and when it exhibits characteristic quantum interference. We have found that the condition for this to happen is that $|U_{11}U_{22}+U_{12}U_{21}|=0$ together with $\rho_{0202}\neq0$ and $\rho_{2020}\neq0$. In general, given a certain redshift $\chi$, specific design of the modes $F_{\omega_0}$ and $F_{\tilde{\omega}_0}$ will change the value of these three key quantities in a desired way. 
The conditions mentioned here can be obtained, for example, by engineering the two modes $F_{\omega_0}$ and $F_{\tilde{\omega}_0}$ to have multiple peaks that alternate \cite{Bruschi:Chatzinotas:2021}. It is also clear that single bell-shaped modes that do not overlap lead immediately to either vanishing $U_{21}$ or $U_{12}$, which therefore implies the destruction of the interference effect.
Experimental detection of this effect would allow us, as mentioned above, to support the validity of quantum field theory in (weakly) curved spacetime, which still lacks experimental corroboration regardless of the many unique and striking theoretical predictions \cite{Birrell:Davies:1982,Carroll:2019}. A promising potential avenue for such tests is the use of Cubesats and other small crafts that are now being considered for use in space-based quantum experiments \cite{Oi:Ling:2017,Joshi:Pneaar:2018,Mazzarella:Lowe:2020}. In this case, small and relatively inexpensive satellites can be deployed at a fraction of the cost of conventional missions, and the craft itself can be potentially loaded with all necessary equipment to perform (reasonable) long-range experiments. One idea can be to use a small collection of such satellites as sources of photons to be detected on Earth \cite{Belenchia:Carlesso:2022}.

Another important aspect that can be explored using the predictions of this work is that of the validity of the Einstein Equivalence Principle (EEP) in a framework where not only gravitational features but also quantum mechanical features of a physical system play a role. The EEP prescribes that the laws of physics reduce to those of special relativity locally (i.e., in regions of spacetime that are small enough) \cite{Carroll:2019}. This is a fundamental statement about Nature, and it is therefore a matter of fundamental interest to know if this principle holds. To date, there are many experiments that have been already performed, and more are planned \cite{Altschul:Bailey:2015,Tino:Cacciapuoti:2020,Bassi:Cacciapuoti:2022}. An even more compelling problem is the validity of the EEP in the quantum domain. While it is implicitly assumed that it does apply, there are different arguments why testing it for free falling quantum systems would be greatly beneficial for our current understanding \cite{Altschul:Bailey:2015}.
We note that this work might provide yet another way to test the EEP, although it does not solve the problem of the EEP for gravitating quantum matter (attempts in this direction already exist \cite{Zych:Brukner:2015}).
Contrary to many proposed and performed experiments, we would not use massive particles (atoms) \cite{Tino:Cacciapuoti:2020}, but massless excitations of a quantum field. Photons can propagate (i.e., ``free fall'') between two users at different height in the gravitational potential, and the shift can be measured using interferometric setups \cite{Scully:Zubairy:1997,Barzel:Bruschi:2022}. Given the high degree of control over photons and the high precisions allowed by photonics, it would be possible to test the universality of the gravitational redshift against, for example, the initial (quantum) state of the photon, the different motion of photons (i.e., varying the trajectory), and the polarization. Since gravitational redshift is to be expected on first principles as a direct consequence of the EEP applied to two accelerated objects that exchange electromagnetic pulses \cite{Carroll:2019}, we conclude that this avenue is yet another dimension that can be explored with the mechanism described here. More work is of course necessary to establish a concrete protocol and put forward a realistic experimental proposal. 

We continue by recalling that novel and advanced theories of Nature predict deviations from those of general relativity that occur in specific (typically high-energy or extremely small scale) regimes. There are several proposals to test different aspects of novel physics in space \cite{Belenchia:Carlesso:2022}. One advantage of the space-based setup is that photons propagating through spacetime might be able to witness deviations from expected kinematics. These effects might be due to, for example, asymmetries as a consequence of anisotropic background spacetimes (effects that can be witnessed by comparing results from experiments with photons propagating in different directions), an ultraviolet cutoff or coarse graining of spacetime among many \cite{Bassi:Cacciapuoti:2022}. In this case, propagation through a long baseline can provide the necessary cumulation of effects that can lead to successful detection. Since the mode-mixing predicted here is a definitive signature of quantum field theory in curved spacetime, any deviation could be amplified in an interferometric-like measurement and therefore detected. We believe that this is another opportunity in support of testing the results of this work.

We also note that mode mixing is a key phenomenon in many area of physics, and it is an ubiquitous operation in quantum optics \cite{Scully:Zubairy:1997}. Neutrino physics is another area where mode mixing has led to a revolutionary new understanding of high-energy physics processes. While previously thought to be massless, neutrinos were subsequently proposed to be massive, a feature that was experimentally confirmed and that requires them to ``mix flavours'' \cite{Bellini:Ludhova:2014,Salas:Forero:2020}. This phenomenon, known as \textit{neutrino oscillations}, can be seen as a form of mode mixing, where three distinct operators (flavours) are mixed unitarily into three new ones  \cite{Bellini:Ludhova:2014}. While we do not present the theory or discuss the implications, we note that neutrinos, like every other realistic particle, will be represented by a wavepkacket of field excitations, which requires updating the mathematical technology developed for ideal sharp-momentum particles in order to take care of all realistic features. Our results can help in addressing some of the issues, including adding the effects of weak gravitational backgrounds on the propagation of the neutrinos as wave-packets. 

Finally, states that exhibit such quantum coherence can be used as resources for quantum computing \cite{Nielsen:Chuang:2010}. It remains an open question how this final aspect can be used constructively in concrete applications.

\vspace{0.9cm}

\section{Conclusion}
We have shown that gravitational redshift cannot be implemented as a unitary operation on the sharp-frequency field-modes alone. Instead, the effects of gravitational redshift on propagating photons can be modelled as a mode-mixer, which shifts excitations from one particular frequency distribution to others. We then showed that this effect can be exploited to induce two-photon Hong-Ou-Mandel-like interference purely as a consequence of the photons propagating in a curved background. This result adds to the existing unique predictions of quantum field theory in curved spacetime. We therefore conclude that our work provides novel insight into the quantum aspects of gravitational redshift \cite{Chang:2018} and, more broadly, the interplay of relativity and quantum mechanics. Experimental verification of this effect is within the reach of near future experimental capabilities.

\acknowledgements
We thank Jan Kohlrus, Claus L\"ammerzahl, Valente Pranubon and Leila Khouri for useful suggestions. The \href{https://pixabay.com/vectors/satellite-artificial-space-data-5174090/}{satellite} of Figure~\ref{fig1} is licensed for free use by \textit{Pixabay}, while the \href{https://pngimg.com/image/25352}{Earth} is licensed for non commercial use under the CC BY-NC 4.0 agreement by \textit{pngimg.com}.

\bibliographystyle{apsrev4-2}
\bibliography{UniRedshift}

\end{document}